\documentclass[aps,showpacs,twocolumn]{revtex4}
\usepackage{epsfig}
\usepackage{amsmath}

\begin{document}

\title{Remarks on the hidden color components in multi-quark study\footnote{Corresponding author: fgwang@chenwang.nju.edu.cn (F. Wang)}}

\author{Fan Wang$^a$, Jialun Ping$^b$, Hongxia Huang$^{b}$}

\affiliation{$^a$Department of Physics, Nanjing University, Nanjing 210093, P.R. China}

\affiliation{$^b$Department of Physics, Nanjing Normal University, Nanjing 210097, P.R. China}

\begin{abstract}
Problems related to the application of hidden color components in multi-quark systems are discussed in this report.
\end{abstract}


\maketitle

\setcounter{totalnumber}{5}

\section{\label{sec:introduction}Introduction}
Hidden color components have been introduced in the study of multi-quark systems since the very beginning
of the quantum chromodynamics (QCD) based nuclear physics. In the study of nucleon-nucleon ($NN$) interaction,
the quark cluster model was introduced in 1978~\cite{wang78}. Very soon in the study of color van der Waals force
the hidden color channel coupling to colorless $NN$ quark cluster channel was invoked~\cite{wang82}. Here
the hidden color channel means the individual nucleon is in color octet state instead of color singlet
but the two color octet nucleons are coupled to an overall color singlet one to fulfil the color confinement
principle. Later on the hidden color channel (component) had been extended to other multi-quark systems where
the individual hadron is colorful but overall color singlet, for example a hidden color penta-quark component
means a color octet baryon with a color octet meson coupling to colorless five-quark state, two individual
color mesons or diquark-antidiquark coupling to a color singlet tetra-quark state, etc.

Harvey discussed the transformation between the six quark cluster bases (also called physical bases) and
the symmetry bases (also called group chain classified bases)~\cite{harvey} in the adiabatic calculation
of $NN$ interaction. He used the $SU^{\tau\sigma}(4)\supset SU^{\tau}(2)\times {SU^{\sigma}(2)}$
group chain classified bases for colorless channels but the $SU^{\tau\sigma}(4)$ for hidden color channels.
(Here we use $\tau$ and $\sigma$ to denote isospin and spin.) Also the phase choice is confused. Chen improved
Harvey's results with a systematic group theory method to calculate the transformation between symmetry bases
and physical bases~\cite{chen}. He pointed out there are 5 hidden color channels for spin-isospin (10), (01)
six-quark state instead of Harvey's one hidden color channel. We extended this transformation from
$SU^{\tau\sigma}(4)\supset SU^{\tau}(2)\times{SU^{\sigma}(2)}$ to
$SU^{f\sigma}(6)\supset (SU^{f}(3)\supset SU^{\tau}(2)\times{U^{Y}(1)})\times SU^{\sigma}(2)$ and
from non-relativistic to relativistic case~\cite{wang95}, here $Y$ means hepercharge $Y=B+S$, $B$ and $S$ are
the baryon number and strangeness. Later on the transformation between physical bases and symmetry bases was
extended to penta-quark systems~\cite{huang07,ping08}.

There have been different applications and understandings of the hidden color components and the transformation
between symmetry bases and physical bases. In this report we will discuss the problems related to the understanding
of these two sets of multi-quark bases and their transformation especially about the physical effects of hidden color components.

\section{Symmetry bases and physical bases and their mutual transformation}
Symmetry and physical bases both were introduced in nuclear physics. Symmetry bases were introduced in
the study of nuclear shell model, where the individual nucleon is assumed to occupy the single particle states
$1s$, $1p$, $2s$, $1d$, etc. to form a fixed many particle configuration. Group theory is used to help construct
the orthonormal many particle bases to facilitate the calculation. In addition to the
$SU^{\tau\sigma}(4)\supset SU^{\tau}(2)\times{SU^{\sigma}(2)}$ spin-isospin part, one usually combines with the
$SU(n)\supset{SO(3)}$ orbital part to form the totally antisymmetric many nucleon bases. In general the $SU(n)$
group is just a mathematical device instead of the real symmetry of the nucleon system. For the ($2s1d$)
configuration, Elliott proposed a physical orbital group chain $SU(3)\supset{SO(3)}$ to describe the rotational
motion of nucleus~\cite{elliott}. Here we show such a symmetry bases (group chain classified bases) as an example,
\begin{eqnarray}
&& (SU^{\tau\sigma}(4)\supset SU^{\tau}(2)\times SU^{\sigma}(2))\times (SU(3)\supset SO(3)), \nonumber \\
&&~~~~~[\tilde{\nu}]~~~~~~~~~IM_I~~~~~~~SM_S~~~~~~~~~~[\nu]~~~~~~~LM_L
\end{eqnarray}
The second line is the irreducible representation (IR) label of $SU^{\tau\sigma}(4)$, $SU^{\tau}(2)$, $SU^{\sigma}(2)$,
$SU(3)$ and $SO(3)$ orbital respectively. These symmetry bases are orthonormal, because the single particle orbital states
are orthonormal.

The physical bases (cluster bases) is introduced in nuclear cluster model. Usually one uses two-cluster bases to describe
the two body scattering or bound state. For example one uses two-alpha cluster bases to study the alpha-alpha scattering.
To call these bases as physical bases is because of they have clear physical meaning. In general the physical basis is expressed as,
\begin{equation}
\Psi_{IM_I}^{SM_S}(A)=\mathcal{A}\{[\psi_{I_1S_1}(A_1)\psi_{I_2S_2}(A_2)]_{IM_I}^{SM_S}F(\mathbf{R})\}, \label{rgm}
\end{equation}
here $\Psi(A)$, $\psi(A_1)$ and $\psi(A_2)$ are internal wave functions for the $A$-nucleon, $A_1$-nucleon and
$A_2$-nucleon systems, respectively, $[~~]_{IM_I}^{SM_S}$ means coupling the individual $I_1$, $I_2$, $S_1$, $S_2$
into the total channel isospin-$I$ and spin-$S$ with the isospin, spin $SU(2)$ Clebsch-Gordan (CG) coefficients, $F(\mathbf{R})$ is the orbital wave function describing the relative motion
between the $A_1$-nucleon cluster and $A_2$-nucleon cluster, $\mathbf{R}_1$, $\mathbf{R}_2$,
$\mathbf{R}=\mathbf{R}_1-\mathbf{R}_2$ are the center of mass (CM) coordinates of $A_1$, $A_2$
nucleon cluster and the relative coordinate between the two clusters. $\mathcal{A}$ is
the unitary antisymmetry operator to antisymetrize the $A$-nucleon wave function. Generally these physical or cluster bases
are not orthogonal ones after antisymmetrization.
There is almost no discussion on the transformation between nuclear symmetry bases and physical bases,
except a general discussion given in our book: \emph{Group Representation Theory for phyicists}~\cite{cpw}.

Quark system has very high symmetry. The group chain $SU^{c\tau\sigma}(12)\supset SU^{c}(3)\times
(SU^{\tau\sigma}(4)\supset SU^{\tau}(2)\times$ $SU^{\sigma}(2))$ is a good symmetry. The even higher
symmetry group chain $SU^{cf\sigma}(18)\supset SU^{c}(3)\times (SU^{f\sigma}(6)\supset
(SU^{f}(3)\supset SU^{\tau}(2)\times U^{Y}(1))\times SU^{\sigma}(2))$ is a good approximate symmetry too.

For single baryon in the three valence quarks model, these group chain classified (symmetry) bases combined
with proper orbital symmetry bases describe the baryon spectroscopy quite well. However, because at the
very beginning of QCD based nuclear physics, the main application of these symmetry quark bases is for
the study of $NN$ interaction, here instead of the one center single particle orbital states used in nuclear
physics and baryon spectroscopy, the two center states are used to construct the orbital symmetry bases.
Usually these two center orbital states are called left center and right center single particle orbit, expressed as,
\begin{eqnarray}
 \phi_l(\mathbf{r}) & = & (\pi b^2)^{-3/4}e^{-(\mathbf{r}+\mathbf{s}/2)^2/{2b^2}}, \nonumber \\
 \phi_r(\mathbf{r}) & = & (\pi b^2)^{-3/4}e^{-(\mathbf{r}-\mathbf{s}/2)^2/{2b^2}}.
\end{eqnarray}
The overlap of these two single particle states is,
\begin{equation}
 \langle\phi_l(\mathbf{r})|\phi_r(\mathbf{r})\rangle=e^{-s^2/{4b^2}}, \label{ov}
\end{equation}
and so these single particle orbits are not orthogonal. Only in the limit of the separation between
two centers $s\rightarrow\infty$, then they become orthogonal. In the separation $s\rightarrow 0$ limit,
the left and right center orbits collapse into the same orbit.

For $NN$ or baryon-baryon ($BB$) scattering study, because one is usually interested in the ground state baryon
scattering, one always assumes each three-quark occupy the same left or right single particle orbit.
Then the six quark will be in the $l^3r^3$ configuration. Based on the out-product rule of the permutation group,
the six quark system can have the following four orbital symmetries [6], [42], [51], [33] of $SU^x(2)$ group consisted
of the left and right orbits. In combination with the color-flavor-spin group chain, one has
the following symmetry group chain for the six-quark systems,
\begin{widetext}
\begin{eqnarray}
& & SU^{xcf\sigma}(36)\supset SU^{x}(2)\times (SU^{cf\sigma}(18)\supset SU^{c}(3)\times (SU^{f\sigma}(6)
    \supset(SU^{f}(3)\supset SU^{\tau}(2)\times U^{Y}(1))\times SU^{\sigma}(2))), \nonumber  \\
& & ~~~[1^6]~~~~~~~~~~~~[\nu]r^3l^3~~~~~~~~~~~~[\tilde{\nu}]~~~~~~~~~[2^3]W~~~~~~~~~[\mu]~~~~~~~~~~~[f]~~~~~
~~~~~~{IM_I}~~~~~~~Y~~~~~~~~~~{SM_S},
\end{eqnarray}
\end{widetext}
here $[1^6]$ denotes totally antisymmetric six-quark state, $[\nu]r^3l^3$, $[\tilde{\nu}]$, $[2^3]W$, $[\mu]$,
$[f]$, $IM_I$, Y, $SM_S$ denote the above discussed $SU(2)$ orbital symmetry and the labels of $SU(18)$ IR,
color $SU(3)$ IR and the Weyl tableau, $SU(6)$ IR, flavor $SU(3)$ IR, isospin $SU(2)$ IR, hypercharge $U(1)$ IR,
spin $SU(2)$ IR respectively. The color part must be in the $[2^3]$ IR due to color singlet requirement,
the Weyl tableau is also fixed. The IR labels of $SU^{x}(2)$ and $SU(18)$ must be conjugate due to totally antisymmetry
condition of multi-quark state.

It should be emphasized that these six-quark symmetry bases are only orthogonal but not normalized because
the single particle orbital states are not orthogonal ones except in the $s\rightarrow\infty$ limit.
In the another limit, $s\rightarrow 0$, the left and right single particle states collapse into the same state.
In this case, all of these symmetry bases go to zero except the totally symmetric one $[\nu]=[6]$ because there is only
one orbital state.

Harvey introduced a symmetry and separation distance $s$ dependent normalization factor $N([\nu],s)$  (Harvey denoted
as $N[f]$)~\cite{harvey}. These normalization factors are,
\begin{eqnarray}
 N([6],s) & = & \sqrt{1+9m^2+9m^4+m^6}~,\nonumber \\
 N([42],s) & = & \sqrt{1-m^2-m^4+m^6}~, \\
 N([51],s) & = & \sqrt{1+3m^2-3m^4-m^6}~, \nonumber \\
 N([33],s) & = & \sqrt{1-3m^2+3m^4-m^6}~.
\end{eqnarray}
The overlap $m=\langle \phi_l(\mathbf{r})|\phi_r(\mathbf{r})\rangle$ is given in Eq.(\ref{ov}). In the
$s\rightarrow\infty$ limit, the overlap goes to zero and so all of these four normalization factors equal 1.
By introducing these normalization factors, all of the four orbital symmetry bases [6], [42], [51], [33] are
orthonormal. In the $s\rightarrow 0$ limit, $m=1$ and so all of these normalization factors go to zero except
the $N([6],s)$. Harvey proved these four orbital bases go to $s^6$, $s^4p^2$; $s^5p^1$, $s^3p^3$ configurations
respectively.

Harvey and Chen both employed the following physical (cluster) bases,
\begin{equation}
\Psi_{IS}(6)=\mathcal{A}[\psi_{I_1S_1}(l^3)\psi_{I_2S_2}(r^3)]_{IS},
\end{equation}
here we use a simplified notation, neglected the intermediate quantum numbers. In fact $\Psi_{IS}(6)$ is overall color singlet
and with fixed $M_I$, $M_S$ and hypercharge (in the three flavor case), the orbital configuration is $l^3r^3$.
The left centered three-quark cluster $\psi_{I_1S_1}(l^3)$ and the right centered one $\psi_{I_2S_2}(r^3)$ are color singlets
($[1^3]$ IR) for colorless channels, and color octet ($[21]$ IR) for hidden color channels, the orbital symmetry is $[3]$,
the IR of $SU^{f\sigma}(6)$ or $SU^{\tau\sigma}(4)$ is restricted to the totally symmetric $[3]$ for color singlet channels,
and in turn the $SU^{f}(3)$ (or $SU^{\tau}(2)$) and $SU^{\sigma}(2)$ are in the $[3]\times[3]$ (flavor decuplet baryons) or
$[21]\times[21]$ (flavor octet baryons) representations. For color octet baryons, The IR of $SU^{f\sigma}(6)$ or
$SU^{\tau\sigma}(4)$ must be $[21]$,  there are $[3]\times[21]$, $[21]\times[21]$ flavor-spin symmetry for spin 1/2 case,
$[21]\times[3]$ for spin 3/2 case, there is an additional $SU^{f}(3)\times SU^{\sigma}(2)$ symmetry $[1^3]\times[21]$
(spin 1/2 flavor singlet $\Lambda$) in the three flavor case. The $[~]_{IS}$ means coupling two three-quark clusters
into isospin $IM_I$, spin $SM_S$ and overall color singlet~\cite{chen}. These symmetry and physical bases include the center of mass motion.

Both symmetry and physical bases form complete orthonormal set if Harvey's normalizations are employed. Every set can be used
to describe the six quark states. The transformation between these two bases sets are unitary transformation. However the
center of mass motion of Harvey's bases are hard to be separated and physical meaning is not clear except for infinite separation. Therefore in the dynamical calculation one prefers to use another physical bases, the resonating group method (RGM) cluster bases as shown in Eq.(\ref{rgm}), which is translational invariant. This physical bases can be rewritten in the form of Harvey and Chen' version. First, one expands the relative motion wave function with Gaussian bases,
\begin{equation}
F(\mathbf{R})=\sum C_i(\frac{3}{2\pi b^2})^{\frac{3}{4}}e^{-\frac{3}{4b^2}}(\mathbf{R}-\mathbf{s}_i)^2,
\end{equation}
second, multiply with a total six-quark center of mass orbital wave function,
\begin{equation}
\Phi(\mathbf{R}_C)=(\frac{6}{\pi b^2})^{\frac{3}{4}}e^{-\frac{3}{b^2}R^2_C},
\end{equation}
then the translational invariant RGM physical six-quark cluster bases can be rewritten in the Harvy and Chen's physical bases
version. The $\mathbf{s}_i$ stands for the center separation parameter in Harvey and Chen's version. It is also the generating
coordinate in the generating coordinate method (GCM). However the Harvey's separation $s$ dependent normalization factor is
impossible to insert. As already emphasized, this physical bases is translational invariant and the center of mass motion is
separated. The weak point of this bases is that {\bf they are not orthonormal}. The transformation between physical bases and
symmetry bases is still as before, but now the {\bf individual transformation coefficient no longer keeps as the probability amplitude}~\cite{ping01}. Another point should be emphasized is the cluster bases are meaningful only if the separation
$s_i$ between the cluster centers are $\geq$ the rms radius of cluster themselves otherwise it is just a mathematical device.

\section{Application of the transformation between symmetry bases and physical bases}

The main application of these bases is in the quark model calculation of hadron-hadron interaction and
multi-quark state. Either physical or symmetry bases are complete and so one can use any set of these bases to do the
calculations. However the most powerful approach is to combine these two sets of bases together in one calculation.
A sophisticated hadron-hadron interaction or multi-quark state calculation usually involved many physical channels.
For example to study the well-known H particle (a $uuddss$ six quark state with hypercharge-isospin-spin $YIJ=000$)
there are 21 physical bases coupling together if the hidden color channels are included~\cite{H}.
To use physical bases to express the multi-quark state has the obvious advantage that the physical meaning of each channel is clear. However to directly
calculate the $21\times 21$ matrix elements with physical bases is cumbersome. Then the transformation between physical
bases and symmetry bases helps to express the matrix elements in terms of the symmetry bases. The advantage in using
the symmetry bases is because of one can use the fractional parentage expansion to reduce the six-body matrix elements
calculation into two-body matrix elements calculation if only two body interactions are included in the Hamiltonian.
We had developed such kind calculation for five and six quark systems~\cite{wang95,huang07,ping08}.

The hidden color channel had been used in the study of color van der Waals force~\cite{wang82}. This study showed that
the un-physical strong color van der Waals force is due to the un-physical use of the color confinement interaction,
which can not be extended to a range much larger than the hadron scale, where the sea quark pair creation should
be taken into account. Indeed the quenched lattice QCD calculation obtained the lineal confinement interaction,
the unquenched lattice QCD obtained a color screening confinement which cut off the un-physical long range color
van der Waals force~\cite{lqcd}.

Another application of hidden color components is in the $NN$ intermediate range attraction. If only quark and gluon
degree of freedom are included in the calculation, one can only obtain the repulsive core of $NN$ interaction.
The hidden color channel coupling combined with quark delocalization helps to obtain the intermediate $NN$ attraction,
which is provided by the $\sigma$ meson exchange both in the meson exchange model and chiral quark model. The quark
delocalization is similar to the electron delocalization (or percolation) in the hydrogen molecule structure.
The quark delocalization and hidden color channel coupling mechanism describes the NN interaction as well as the meson exchange models and explains the similarity between nuclear force and molecular force naturally, it shows the $NN$ interaction is a QCD duplication of the QED molecular force~\cite{wang92,chen07,huang11}.

Brodsky, Ji and others discussed various effects of hidden color channels in multi-quark systems~\cite{jicr}.

First, they try to attribute the repulsive core of short range $NN$ interaction to the effect of hidden color component.
This is questionable. The NN $IJ=01$ deuteron, $\Delta\Delta$ $IJ=01,03$ channels have the same amount of hidden color components (see the following TABLE II,III), only the deuteron channel has the short range repulsion, the other two $\Delta\Delta$ channels both have short range attraction. In fact the short range repulsion of $NN$ channel and the attraction of the $\Delta\Delta$ channel both are due to the color magnetic interaction (CMI).
CMI contributes attraction to the internal energy of an octet baryon, $\langle CMI \rangle_N=-8C$ (see Eq.(\ref{CMIN}) below),
because there are equal attractive and repulsive quark-quark pairs within octet baryon.
On the contrary there are only repulsive $qq$ pairs within decuplet baryon, $\langle CMI \rangle_{\Delta}=8C$ (sea Eq.(\ref{CMID}) and
this causes the decuplet baryon about 300 MeV massive than the octet baryon.
When two nucleons with deuteron quantum number $IJ^P=01^+$ merge into an orbital totally symmetric color singlet six-quark state (denoted as $d$), there are more repulsive
quark-quark pairs than the attractive ones (Eq.(\ref{diffN})), and when two decuplet baryons with the same quantum number $IJ^P=01^+$ merge into orbital
totally symmetric color singlet six-quark state (it is same as the state merged from two nucleons if the six quarks
stay at the same orbital state) there are attractive
quark-quark pairs and cause the effective attraction between decuplet baryons (Eq. (\ref{diffD})). For reference,
the coupling of two $\Delta$'s to $IJ^P=03^+$ (denoted as $d^*$) (Eq. (\ref{diffD*})) are also shown below.
\begin{widetext}
\begin{eqnarray}
\langle CMI \rangle_N & = & -3C \langle \boldsymbol{\lambda}_2\cdot \boldsymbol{\lambda}_3 \rangle_A
\left[  \langle \boldsymbol{\sigma}_2\cdot \boldsymbol{\sigma}_3 \rangle_A  +
  \langle \boldsymbol{\sigma}_2\cdot \boldsymbol{\sigma}_3 \rangle_S  \right]/2  =¡¡ -8C, \label{CMIN} \\
\langle CMI \rangle_{\Delta} & = & -3C \langle \boldsymbol{\lambda}_2\cdot \boldsymbol{\lambda}_3 \rangle_A
  \langle \boldsymbol{\sigma}_2\cdot \boldsymbol{\sigma}_3 \rangle_S  =¡¡8C, \label{CMID} \\
\langle CMI \rangle_d & = & -15C \left\{\langle \boldsymbol{\lambda}_5\cdot \boldsymbol{\lambda}_6 \rangle_A
 \left[ \frac{5}{30}\langle \boldsymbol{\sigma}_5\cdot \boldsymbol{\sigma}_6 \rangle_A  +
 \frac{13}{30} \langle \boldsymbol{\sigma}_5\cdot \boldsymbol{\sigma}_6 \rangle_S  \right]+
 \langle \boldsymbol{\lambda}_5\cdot \boldsymbol{\lambda}_6 \rangle_S
 \left[ \frac{5}{30}\langle \boldsymbol{\sigma}_5\cdot \boldsymbol{\sigma}_6 \rangle_A  +
 \frac{7}{30} \langle \boldsymbol{\sigma}_5\cdot \boldsymbol{\sigma}_6 \rangle_S  \right] \right\} \nonumber \\
  &  = & \frac{8}{3} C, \\
\langle CMI \rangle_{d^*} & = & -15C \left( \frac{3}{5}\langle \boldsymbol{\lambda}_5\cdot \boldsymbol{\lambda}_6 \rangle_A
 \langle \boldsymbol{\sigma}_5\cdot \boldsymbol{\sigma}_6 \rangle_S  +\frac{2}{5}\langle \boldsymbol{\lambda}_5\cdot
 \boldsymbol{\lambda}_6 \rangle_S \langle \boldsymbol{\sigma}_5\cdot \boldsymbol{\sigma}_6 \rangle_S  \right)  =  16 C, \\
& & \hspace{-0.8in} \langle CMI \rangle_d -2 \langle CMI \rangle_N = \frac{56}{3}C, \label{diffN} \\
& & \hspace{-0.8in} \langle CMI \rangle_d -2 \langle CMI \rangle_{\Delta} = -\frac{40}{3}C, \label{diffD} \\
& & \hspace{-0.8in} \langle CMI \rangle_{d^*} -2 \langle CMI \rangle_{\Delta} = 0, \label{diffD*}
\end{eqnarray}
\end{widetext}
where $C$ is the orbital matrix element, the subscripts $A$ and $S$ denote antisymmetric and symmetric.
The hidden color components couple to
these octet and decuplet baryon-baryon channels equally and this shows the $NN$ repulsive core is not due to
the effect of hidden color channels~\cite{chen}.
\begin{table}
\caption{The color-spin part of the matrix elements of CMI with totally symmetric orbits (unit: C)}
\begin{tabular}{cccc} \\ \hline
$[\nu]IJ^P$ & $[3]\frac{1}{2}\frac{1}{2}^+$  & $[3]\frac{3}{2}\frac{3}{2}^+$ & $[6]01^+$  \\ \hline
 M.E. &  $-8$  &  $8$  & $8/3$     \\ \hline
$[\nu]IJ^P$  & ~$[6]01^+-2([3]\frac{1}{2}\frac{1}{2}^+)$~  & ~~$[6]01^+-2([3]\frac{3}{2}\frac{3}{2}^+)$ & \\ \hline
 M.E. &   $56/3 $  & $-40/3 $ & \\ \hline
$[\nu]IJ^P$  & $[6]03^+$ & ~$[6]03^+-2([3]\frac{3}{2}\frac{3}{2}^+)$~  & \\ \hline
 M.E. &  16 & $0 $  & \\ \hline
\end{tabular}
\end{table}

They also try to attribute hadron interaction to the color Van der Waals force due to multi-gluon exchange
between colorless hadron through the intermediate hidden color components. R. Machleidt also said in ~\cite{mach}: {\it the force between nucleons is a residual color interaction similar to the van der Waals force between neutral molecules.} In fact, the interaction
between two colorless nucleons is quite similar to that between two electro neutral atoms. The molecular covalence bond
or molecular force and electro van der Waals interaction between two electro neutral atoms had been studied
thoroughly~\cite{schiff}.
When two hydrogen atoms are close
together the electrons originally localized in separate atom mutually delocalize to other atom and distort
the electron orbit, (when two electrons spin is anti-parallel) the charge density of electrons in between
two protons is a little higher than other regions and resulted in the attraction to bind the two atoms into
hydrogen molecule. It can be approximated by the Morse potential. It has a repulsive core and an intermediate
range attraction then a decreasing attraction tail. The electro van der Waals attraction of the form
$-6.5\,e^2a^2/R^6$ ($a$ is the hydrogen atomic Bohr radius and $R$ is the separation of two atomic center of mass)
is a long range tail adding to the Morse potential. Nuclear force (take the deuteron channel as example)
has quite the same form except the length and energy scale difference (see Fig.~1). As explained in the standard model
chart made by the US Contemporary Physics Education Project (CPEP): {\it The strong binding of color-neutral
protons and neutrons to form nuclei is due to residual strong interactions between their color charged constituents.
It is similar to the residual electrical interaction that binds electrical neutral atoms to form molecules.}
The quark delocalization color screening model realizes such a mechanism and explains why the nuclear force and
molecular force are similar, describes the NN interaction as well as the meson exchange with much less parameters.~\cite{wang92}. We also calculated the hidden color channels coupling to colorless $NN$ channel,
it is also similar to the molecular case these hidden color channels coupling gives rise the color van der Waals
interaction which enhances the $NN$ attraction a little bit but does not play the dominant role in the $NN$
interaction~\cite{huang11}. Therefore we suggest to differentiate the nuclear force from the color
van der Waals force even though they contribute to the $NN$ interaction.

\begin{figure*}
\epsfxsize=6.0in \epsfbox{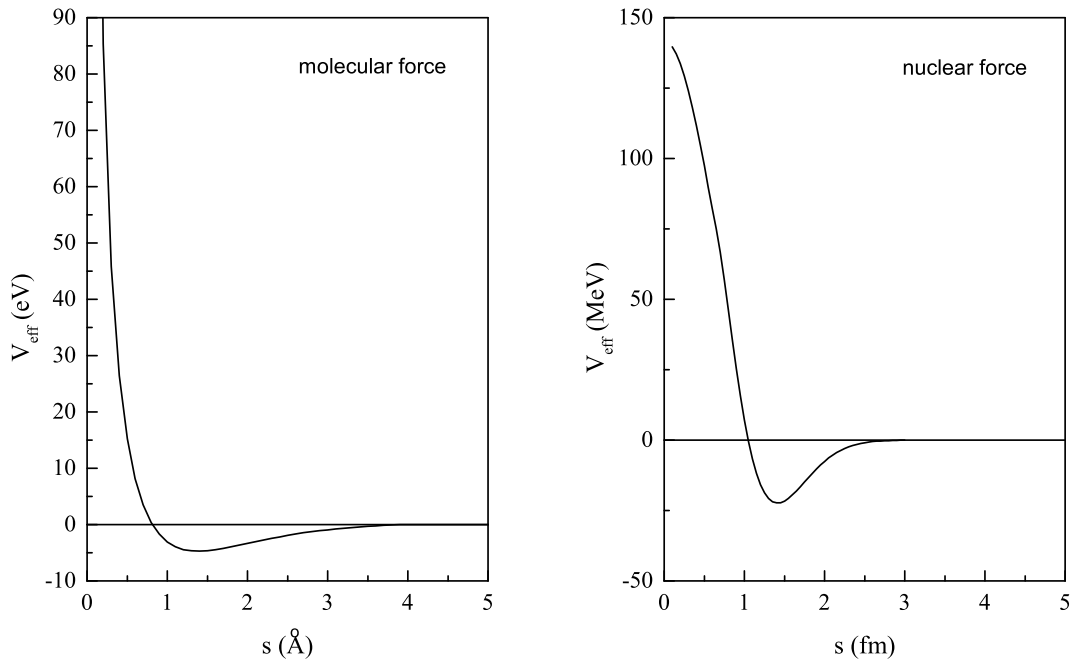}
\vspace{-0.1in}
\caption{Similarity of molecular force and nuclear force}
\end{figure*}
\begin{table*}[ht]
\caption{The transformation between the physical bases and symmetry bases for $YIJ=201$.
The column heads are the orbital $SU_2^x$ and isospin-spin $SU^{\tau\sigma}_4$ symmetries, $[\nu][\mu]$.}
\begin{tabular}{cccccccc} \hline
                & ~~$[6][33]$~~ & ~~$[51][321]$~~ & ~~$[42][51]$~~ & ~~$[42][411]$~~ & ~~$[42][33]$~~ & ~~$[42][321]$~~
                & ~~$[42][2211]$~~ \\ \hline
 $NN$           & $\sqrt{1/9}$   &  0  & $\sqrt{4/9}$ & $\sqrt{4/9}$ &  0 &  0 &  0 \\
 $\Delta\Delta$ & $-\sqrt{4/45}$ &  0  & $\sqrt{5/9}$   & $-\sqrt{16/45}$ &  0 &  0 &  0 \\
 ~~$\Delta_8^*\Delta_8^*$~~ & $\sqrt{1/9}$ &  0  & 0 & $-\sqrt{1/36}$ & $-\sqrt{1/18}$ & $\sqrt{16/45}$ & $\sqrt{9/20}$ \\
 $N_8N_8$ & $\sqrt{2/9}$ &  0  &  0  & $-\sqrt{1/18}$ & $\sqrt{4/9}$  & $-\sqrt{8/45}$  &  $\sqrt{1/10}$ \\
 $N_8N_8^*$ & $-\sqrt{2/9}$ &  $\sqrt{1/2}$ &  0  & $\sqrt{1/18}$  & $\sqrt{1/9}$ & $\sqrt{1/90}$ & $\sqrt{1/10}$ \\
 $N_8^*N_8$ & $\sqrt{2/9}$ & $\sqrt{1/2}$  & 0 & $-\sqrt{1/18}$ & $-\sqrt{1/9}$  & $-\sqrt{1/90}$ & $-\sqrt{1/10}$ \\
 $N^*_8N^*_8$ & $\sqrt{1/45}$  & 0  &  0  &  $-\sqrt{1/180}$ &  $\sqrt{5/18}$ & $\sqrt{4/9}$ & $-\sqrt{1/4}$ \\ \hline
\end{tabular}
\label{pts01}
\end{table*}

To emphasize the hidden color channel effect, Ji claimed the number of hidden color channels
increases rapidly as the quark number increasing: For three-quark system, there is only one colorless state.
For six-quark system, this number increases to five with one color singlet channel and four hidden color
channels. For nine-quark system, there are one color singlet channel and 41 hidden color channels.
Here they confused the Young tableaux of permutation group with the colorless states of color $SU_3$ group which is labeled
by Weyl tableaux. All the color singlet states, no matter how many quarks are included, there is only one
Weyl tableau and it is the one dimensional IR of the color $SU(3)$ group. A physical color
singlet quark state must be totally antisymmetric, one has to use CG coefficients of the permutation group
to couple the whole Young tableaux of a Young diagram $[k^3]$ ($k=n/3$, $n$ is the number of quarks) for color part
to the whole Young tableaux of
the conjugate Young diagram $[3^k]$ for orbital-flavor-spin part to form the totally antisymmetric one.
\begin{equation}
\Psi_{[1^n]}^{cxf\sigma}=\sum_{m_1,m_2} C_{[k^3]m_1,[3^k]m_2}^{[1^n]1} \chi^c_{[k^3]m_1}\psi^{xf\sigma}_{[3^k]m_2}.
\end{equation}
As for how many states (color singlet and hidden color states) of a multi-quark system, it depends on all degrees
of freedom. For orbital ground ones, $[n]$, if one restrict to $SU^{\tau\sigma}(4)\supset SU^{\tau}(2)\times SU^{\sigma}(2)$ ,
then even for 3-quark system (the color $SU^c(3)$ and permutation $S_3$ group both are 1-dim representations), there are
still two isospin-spin $1/2$ states (proton and neutron) and four isospin-spin $3/2$ $\Delta$ states (different
spin projection has not been counted here). If one allows orbital excitation, there are in principle infinite
baryon excited states. For 6-quark system, still restricting to $SU^{\tau\sigma}(4)\supset SU^{\tau}(2)\times SU^{\sigma}(2)$
case, even we further restrict to allow only two orbital states and deuteron quantum number $IJ=01$ we still have
two color singlet $NN$ and $\Delta\Delta$ channels and 5 hidden color channels~\cite{chen} not just one color singlet
and 4 hidden color channels as claimed in Ref.\cite{jicr} (see Table \ref{pts01}). On the other hand if we restrict to
$s^6$ configuration as mentioned in Ref.\cite{jicr}, then there will be only one state with deuteron quantum numbers
$YIJ=201$. In this case, the only possible orbital symmetry is $[6]$ of $U^x(1)$, and the color symmetry is physically
restricted to $[222]$ of $SU^c(3)$, so the isospin-spin symmetry must be $[33]$ of $SU^{\tau\sigma}(4)$, and
only one symmetric basis is obtained (different isospin and spin projection has not been counted here).
correspondingly, all the 7 physical bases are collapsed to one.
\begin{table}[ht]
\caption{The transformation between the physical bases and symmetry bases for $YIJ=203$.}
\begin{tabular}{ccc} \hline
    &  $~~~~~[6][33]~~~~~$ & $~~~~~[42][33]~~~~~$ \\ \hline
 $~~~~\Delta\Delta~~~~$   & $\sqrt{1/5}$ &  $\sqrt{4/5}$ \\
 $CC$   & $\sqrt{4/5}$ &  $-\sqrt{1/5}$ \\ \hline
\end{tabular}
\label{pts}
\end{table}

The long expected multi-quark state, the ``inevitable dibaryon $d^*$"~\cite{goldman} had been confirmed by the WASA-at-COSY
collaboration in 2014~\cite{clement} through many years kinematical complete measurements and the polarized $np$ scattering.
The measured resonance mass is 2.37 GeV and the width is about 70 MeV. The quantum numbers is fixed to be $IJ^P=03^+$.
The partial width of resonating scattering is about 12 MeV~\cite{bashkanov}, which is consistent with our Feshbach
resonance scattering calculated result 14 MeV~\cite{ping09}, which shows that the $d^*$ dibaryon is quite possible through
the intermediate di-$\Delta$ production. This theoretical point is consistent with the WASA-at-COSY measurements.
However the total width 70 MeV is much smaller than the sum of two free $\Delta$s' width, also smaller than the
binding reduced two-$\Delta$ width~\cite{ping08}, which might means due to the strong attraction between two $\Delta$s,
it shrinks into a compact six-quark state immediately. Our quark model calculated rms radius of $d^*$ is about 0.9 fm,
as small as a nucleon~\cite{NPA688}. Bashkanov, Brodsky and Clement try to understand this small width as due to the dominant
hidden color components of the compact $d^*$~\cite{jicr}. They use the transformation table between symmetry bases and
physical bases given by Harvey~\cite{harvey} (see Table \ref{pts}),
and assigned the $d^*$ resonance as the totally symmetric orbital $s^6$ states. So 80\% of the $d^*$ state is
in the hidden color component which reduced the $d^*$ decay width. This explanation has the following problems:
(1) If all the six quarks concentrate in the $s^6$ configuration, the orbital part of the six-quark state must be in the
totally symmetric IR [6] of the symmetric group $S_6$, the isospin-spin part must be in the [33] (isospin [33]
and spin [6]) IR of $SU^{\tau\sigma}(4)\supset SU^{\tau}(2)\times SU^{\sigma}(2)$ and so there is only one totally
antisymmetric six-quark state,
\begin{equation}
|[6]s^6[222]W_c,[33][33]I=0,[6]J=3\rangle.
\end{equation}
Correspondingly there is only one physical (or cluster) basis, the colorless $\Delta$-$\Delta$ component and the
hidden color $CC$ component collapse into the same one after the antisymmetrization. The overlap of these
two physical bases at the different separations between two clusters are shown in Table \ref{overlap}.
When the two clusters are well separated ($s\rightarrow \infty$), the two physical bases, colorless
$\Delta$-$\Delta$ and hidden color $CC$, are orthogonal. Nevertheless, when the two clusters are merged into
one cluster ($s\rightarrow 0$), the two physical bases are the same.
\begin{table}
\caption{The overlap of two physical bases in different separations with quantum numbers $YIJ^P=203^+$.}
\begin{tabular}{ccccccccc} \hline
$S$ (fm) & 0.1  & 0.2  &  0.3  & 0.4 & 0.5  & ~1.0~  & 2.0  & ~3.0~ \\ \hline
overlap & 0.99996 & 0.999 & 0.997 & 0.99 & 0.98 & 0.7  & 0.007  & 0 \\ \hline
\end{tabular}
\label{overlap}
\end{table}

The Harvey's table shown above is based on the assumption that there are two orbital states, the left centered one $l$
and the right centered one $r$, the six-quark is in the configuration $l^3r^3$ and there are two orbital symmetries
$[6]l^3r^3$ and $[42]l^3r^3$ for six-quark symmetry bases. Harvey uses a limiting process to obtain the separating
$s_i\rightarrow 0$ limit. Using Harvey's symmetry bases, an adiabatic calculation of the $YIJ=203$ system was performed
and the results are shown in Table \ref{h203}. From the table, we can see that the $\Delta\Delta$ channel has a
similar energy with the hidden-color channel when the separation between two clusters are small, and the two
channels contribute almost the same to the eigenstates. If the separation becomes large, the $\Delta\Delta$ channel
has smaller energy than that of hidden-color channel, and it is the main component of the eigenstate with lower energy.
It is expected that the dynamical calculation will confirm that the main component of $d^*$ state is still the
$\Delta\Delta$ channel, although it is very difficult to do the dynamical calculation using Harvey's symmetry bases.
\begin{table}[ht]
\caption{The energies of $\Delta\Delta$, hidden-color channels and eigenstates in adiabatic approximation
for $YIJ=203$ system using Harvey's symmetry bases (unit, MeV).}
\begin{tabular}{cccccccc} \hline
 $S$ (fm)   & ~~$E_{\Delta\Delta}$~~ & ~~$E_{cc}$~~ & ~~$E_{eigen}$~~  \\ \hline
  0.1       &  2976                  &   2919       &   2630       \\
  0.3       &  2928                  &   2888       &   2590       \\
  0.5       &  2838                  &   2834       &   2519       \\
  0.7       &  2724                  &   2775       &   2430       \\
  0.9       &  2604                  &   2739       &   2347       \\
  1.1       &  2503                  &   2785       &   2308       \\
  1.3       &  2441                  &   3000       &   2338       \\
  1.5       &  2427                  &   3432       &   2398       \\
  1.7       &  2439                  &   3973       &   2435       \\
  2.0       &  2456                  &   4754       &   2455       \\
  2.5       &  2463                  &   6084       &   2463       \\
  3.0       &  2464                  &   7654       &   2464       \\  \hline
\end{tabular}
\label{h203}
\end{table}

If the six quarks are compacted in one center, then one has to use the single center shell
model orbits to construct the multi-body symmetry bases and the cluster bases as we explained in our group theory
book~\cite{cpw}. One has to construct both symmetry bases and cluster bases from fixed configuration, such as
$s^6$, $s^4p^2$, etc. For $s^6$ we have explained before, for $d^*$ quantum number $IJ^P=03^+$ there is only one
symmetry basis and also only one cluster basis (neglect the different spin orientations). For $s^4p^2$ configuration
one has to use the configuration mixing shell model approach, i.e., the $SU(1+3)\supset(U(1)\times SU(3)$ group chain
to construct the six-quark orbital symmetry bases and combined with the $SU^{cf\sigma}(18)\supset SU^{c}(3)\times
(SU^{f\sigma}(6)\supset SU^{f}(3)\supset(SU^{\tau}(2)\times U^{Y}(1))\times SU^{\sigma}(2))$ to obtain the totally
antisymmetric six-quark symmetry bases, then use the isoscalar factors to obtain the transformation between
symmetry bases and cluster bases. The results are listed in Table \ref{su3+1}. When the orbital symmetry of the
three-quark cluster is limited to be $[3]$, Harvey's table is reproduced. However, there are other
three-quark clusters for the configuration $s^4p^2$, and the lowest state should be one with the 
configuration $s^6$, the physical state should be the combination all bases.
\begin{table}[ht]
\renewcommand\arraystretch{1.5}
\caption{The transformation between the physical bases and symmetry bases for $SU(3+1)$
with configuration $s^4p^2$ and $YIJ$=203. The heads of the columns are $[\tilde{\nu}][f]$. 
The correspondence between symbols and symmetries
$[\nu][c][f][\sigma]$ (orbital, color, flavor and spin) for three-quark clusters are:
$N^*$: [21][111][21][3]; $N^*_8$: [111][21][21][3]; $N_8^{\prime *}$: [21][21][21][3];
$N_8^{\prime \prime *}$: [3][21][21][3]; $\Delta$: [3][111][3][3];
$\Delta_8$: [21][21][3][3]. $\overline{xy}=(xy+yx)/\sqrt{2}$, $\widetilde{xy}=(xy-yx)/\sqrt{2}$.}
\begin{tabular}{cc}
\begin{tabular}{c}
\begin{tabular}{cc} \hline
                & $[33][33]$ \\ \hline
 $N_8^*N_8^*$   & 1   \\ \hline
\end{tabular} \\
\begin{tabular}{cc} \hline
                                        & $[411][33]$ \\ \hline
 $\widetilde{~~N_8^*N_8^{\prime *}~~}$  & 1   \\ \hline
\end{tabular} \\
\end{tabular} &
\begin{tabular}{cccc} \hline
                       & $[411][33]$ & $[33][33]$  & $[2211][33]$ \\ \hline
 $\Delta_8 \Delta_8$   & $-\sqrt{1/2}$  & $\sqrt{1/4}$  & $\sqrt{1/4}$ \\
 $N_8^{\prime *}N_8^{\prime *}$   & 0  & $-\sqrt{1/2}$  & $\sqrt{1/2}$ \\
 $N^* N^*$   & $\sqrt{1/2}$  & $\sqrt{1/4}$  & $\sqrt{1/4}$ \\ \hline
\end{tabular} \\
\end{tabular}

~~

\begin{tabular}{cc}
\begin{tabular}{c}
\begin{tabular}{cc} \hline
                            & $[411][33]$ \\ \hline
 $\widetilde{~~N_8^*N_8^{\prime\prime *}~~}$   &  $-1$   \\ \hline
\end{tabular} \\
\begin{tabular}{cc} \hline
                            & $[2211][33]$ \\ \hline
 $\overline{N_8^{\prime *}N_8^{\prime\prime *}}$   &  $-1$   \\ \hline
\end{tabular} \\
\end{tabular} &
\begin{tabular}{ccc} \hline
                            & $[2211][33]$ & $[1^6][33]$   \\ \hline
 $N_8^{\prime\prime *}N_8^{\prime\prime *}$   & $\sqrt{1/5}$  & $-\sqrt{4/5}$ \\
 $\Delta \Delta$   &  $-\sqrt{4/5}$  & $-\sqrt{1/5}$ \\ \hline
\end{tabular} \\
\end{tabular}
\label{su3+1}
\end{table}

Harvey's transformation in the $S_i\rightarrow 0$ limiting case mixes the $s^6$ configuration and the $s^4p^2 $
configuration is very special. We like to emphasize again that the transformation between symmetry and physical bases is restricted in the same one configuration. On the other hand, if the six quarks all compact in a very small space, still using
the cluster bases is not physically meaningful. It is just a six-quark single shell model states. The color structure
is no longer limited to colorless quark cluster or hidden color quark cluster, but also three di-quark cluster,
even a quark benzene structure,
\begin{figure}
\epsfxsize=4.0in \epsfbox{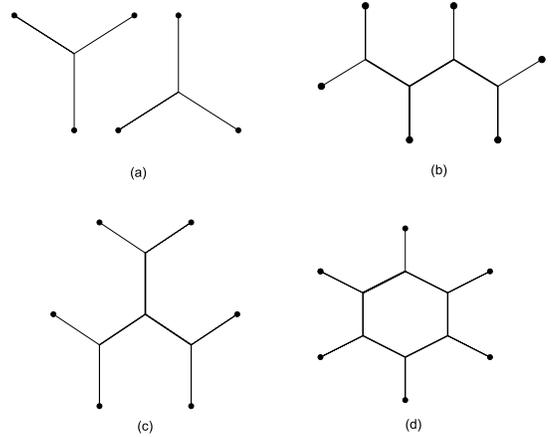}
\vspace{-0.1in}
\caption{The color structures of six-quark system}
\end{figure}

Gonzales and Vento criticized the hidden color is a spurious concept~\cite{GV86}. Their argue is based on the fact that the
hidden color components of a multi-quark system can be replaced by the colorless ones through recoupling.
Mathematically it is true for any colorless $q^n{\bar{q}}^m$ system, $n-m=3k,~ k$ is an integer, one can always find
the colorless hadron bases which are a complete set~\cite{wang85, wang89}. However this does not mean hidden color is not physical ones.
First, if there is gluon excitation, the hidden color meson (baryon) component is unavoidable to form the hybrid
meson (baryon). Even for the pure multi-quark system, the hidden color component is still a physical object.
It can be replaced by the colorless hadron degree of freedom is due to the present quark description is not complete.
A colorless object belongs the color singlet representation. It must be local gauge invariant under color $SU^c(3)$
gauge transformation. The present quark description neglect the gauge link which is needed to guarantee the
multi-quark state local gauge invariant.  If we take into account the different interaction strength for
different color couplings, such as what has been suggested from lattice QCD string structures, the hidden color
is no longer just a coupling scheme. In the color-string quark model the different strengths of different color strings have been included in the model Hamiltonian~\cite{deng08}. Obviously the hidden color components can not be eliminated any more.
The production of multi-quark states from electron-positron collider will
be able to measure the fragmentation function, which might give color structure information immediately
before hadronization.

\section{Summary}

Quark and gluon are colorful object. Due to color confinement there is no colorful observable. However, overall colorless but individual ones colorful, the hidden color component, is possible. It is a new degree of freedom of multi-quark system and nuclei. The present description only describes the $SU^c(3)$ coupling and so for the pure quark-antiquark systems the hidden color component can be replaced by colorless hadron component through quark rearrangement. This is due to the present description neglected the color $SU^c(3)$ local gauge invariant condition which calls for gauge link to link different color objects in different space-time. If the gauge link is incorporated in the description of a multi-quark system then the hidden color component no longer to be replaced by colorless components. For quark-gluon hybrids the hidden color components is unavoidable.

The physical effects of hidden color components are interesting and will be a new facet of QCD. By the end of 1970s the 
color van der Waals force had been a hot topic. However, because we don't have a thorough understanding of confinement we
have not a thorough understanding of color van der Waals force as the QED case yet. Feinberg and Sucher analyzed the 
limit of long range hadron interaction~\cite{FS}. We use a colorless-hidden color channel coupling studied the 
van der Waals force and found the phenomenological confinement interaction will lead to a too strong long-range $NN$ 
interaction which violate the experimental limit obtained by Feinberg and Sucher~\cite{FS,WH84}. We also pointed out 
the linear or quadratic confinement
will be screened by the $q\bar{q}$ excitation and this had been confirmed by the lattice QCD calculation~\cite{lqcd}.

Brodsky and Ji studied various physical effects of hidden color components, from the short range repulsive 
core of $NN$ interaction to the small width of the $d^*$ dibaryon. We have analyzed in the text that the hidden color 
components do not play critical role in these physical effects. In all of these applications,
hidden color components is just a cluster approximation. If the multi-quark system is a compact one such as the $d^*$ 
dibaryon and the short range $NN$ system, the cluster model is no longer adequate. The color structure will have more 
varieties, not just the colorless and hidden color components as described in the quark cluster model. One has to 
deal with the many body problem directly.

Hidden color components is a new degree of freedom and will have physical effects for multi-quark systems 
which should be studied further.

\acknowledgments
The work is supported partly by the National Natural Science Foundation of China under Grant 
Nos. 13001026, 11535005, 11775118 and 11675080.

\end{document}